\documentclass[aps,prb,twocolumn,superscriptaddress,showpacs]{revtex4}

\usepackage{graphicx}
\usepackage{dcolumn}
\usepackage{bm}

\begin{document}

\title{Doping dependence of phonon and quasiparticle heat transport
of pure and Dy-doped Bi$_2$Sr$_2$CaCu$_2$O$_{8+\delta}$ single
crystals}

\author{X. F. Sun}
\email[]{xfsun@ustc.edu.cn}

\affiliation{Hefei National Laboratory for Physical Sciences at
the Microscale, University of Science and Technology of China,
Hefei, Anhui 230026, P. R. China}

\affiliation{Central Research Institute of Electric Power
Industry, Komae, Tokyo 201-8511, Japan}

\author{S. Ono}
\affiliation{Central Research Institute of Electric Power
Industry, Komae, Tokyo 201-8511, Japan}

\author{X. Zhao}
\affiliation{Department of Astronomy and Applied Physics,
University of Science and Technology of China, Hefei, Anhui
230026, P. R. China}

\author{Z. Q. Pang}
\affiliation{Hefei National Laboratory for Physical Sciences at
the Microscale, University of Science and Technology of China,
Hefei, Anhui 230026, P. R. China}

\author{Yasushi Abe}
\affiliation{Central Research Institute of Electric Power
Industry, Komae, Tokyo 201-8511, Japan}

\author{Yoichi Ando}
\affiliation{Institute of Scientific and Industrial Research,
Osaka University, Ibaraki, Osaka 567-0047, Japan}

\date{\today}

\begin{abstract}

The temperature and magnetic-field ($H$) dependences of thermal
conductivity ($\kappa$) of Bi$_2$Sr$_2$CaCu$_2$O$_{8+\delta}$
(Bi2212) are systematically measured for a broad doping range by
using both pure Bi2212 single crystals with tuned oxygen contents
and Bi$_2$Sr$_2$Ca$_{1-x}$Dy$_x$Cu$_2$O$_{8+\delta}$ (Dy-Bi2212)
single crystals with different Dy contents $x$. In the underdoped
samples, the quasiparticle (QP) peak below $T_c$ is strongly
suppressed, indicating strong QP scattering by impurities or
oxygen defects, whereas the phonon conductivity is enhanced in
moderately Dy-doped samples and a phonon peak at 10 K is observed
for the first time in Bi2212 system, which means Dy$^{3+}$ ions
not only introduce the impurities or point defects but also
stabilize the crystal lattice. The subkelvin data show that the QP
heat conductivity gradually decreases upon lowering the hole
doping level. The magnetic-field dependence of $\kappa$ at
temperature above 5 K is mainly due to the QP scattering off
vortices. While the underdoped pure Bi2212 show very weak field
dependence of $\kappa$, the Dy-doped samples present an additional
``dip"-like term of $\kappa(H)$ at low field, which is discussed
to be related to the phonon scattering by free spins of Dy$^{3+}$
ions. For non-superconducting Dy-Bi2212 samples with $x \simeq$
0.50, an interesting ``plateau" feature shows up in the low-$T$
$\kappa(H)$ isotherms with characteristic field at 1 -- 2 T, for
which we discuss the possible revlevance of magnon excitations.

\end{abstract}


\maketitle

\section{Introduction}

Low-temperature heat transport has been extensively studied for
high-$T_c$ cuprates, since it not only provides the most
straightforward information of quasiparticle (QP) transport
properties and the peculiar electronic state, but also shows rich
physics like the magnon heat transport of low-dimensional spin
system, the phonon heat transport and its interaction with
peculiar spin/charge order.

It was predicted that the heat transport of nodal QPs in a
$d$-wave superconductor presents a ``universal"
behavior,\cite{Hussey_Review, Graf, Durst} that is, the residual
conductivity in the zero-$T$ limit $\kappa_0/T$ is independent of
QP scattering rate. Physically, it results from the exact
compensation between the decrease of scattering time and the
increase of zero-energy density of states with increasing impurity
density. In the standard self-consistent $T$-matrix approximation
(SCTMA) theory,\cite{Hussey_Review} when the impurity bandwidth
$\gamma$ is in the universal limit $k_BT \ll \gamma \ll \Delta_0$
($\Delta_0$ is the maximum gap), $\kappa_0/T$ is expressed as
\cite{Graf, Durst}
\begin{equation}
\frac{\kappa_0}{T} = \frac{k_B^2}{3 \hbar} \frac{n}{c} \left(
\frac{v_F}{v_2} + \frac{v_2}{v_F} \right) \simeq \frac{k_B^2}{3
\hbar} \frac{n}{c} \frac{v_F}{v_2}, \label{k0/T}
\end{equation}
with $n$ the number of $\rm{CuO_2}$ planes per unit cell, $c$ the
$c$-axis lattice constant, and $v_F$ ($v_2$) the QP velocity
normal (tangential) to the Fermi surface at the gap node.

The early measurements on nearly optimally-doped
YBa$_2$Cu$_3$O$_y$ (YBCO) \cite{Taillefer} and
Bi$_2$Sr$_2$CaCu$_2$O$_{8+\delta}$ (Bi2212) \cite{Behnia, Chiao}
with various impurity scattering strengths were found to be
supportive of the ``universal" behavior. If valid, the universal
nature is very useful because one can obtain $\Delta_0$ from the
bulk measurement of $\kappa$.\cite{Sutherland_PRB, Hawthorn_Tl}
Sutherland {\it et al.} \cite{Sutherland_PRB} analyzed their
residual conductivity data of YBCO in the framework of the
``universal" thermal conductivity and obtained the gap size
$\Delta_0$ from $\kappa_0/T$; based on this analysis, it was
argued that the $\Delta_0$ values coincide with those obtained
from other spectroscopic measurements. Together with the similar
analysis on the $\kappa$ data of Tl$_2$Ba$_2$CuO$_{6+\delta}$
(Tl2201) in a wide overdoping range,\cite{Hawthorn_Tl} a pure and
simple $d$-wave superconducting state throughout the phase diagram
was proposed. However, as more detailed and accurate data are
accumulated, it has become obvious that there are cases that do
not seem to obey the ``universal" behavior, including the increase
of $\kappa_0/T$ with doping level confirmed in
La$_{2-x}$Sr$_x$CuO$_4$ (LSCO),\cite{Takeya, Sutherland_PRB}
YBCO,\cite{Sutherland_PRB, Sun_YBCO, Sun_logT} and
Bi$_2$Sr$_{2-x}$La$_x$CuO$_{6+\delta}$ (BSLCO),\cite{Ando_BSLCO}
and the varnishing residual thermal conductivity in an underdoped
YBa$_2$Cu$_4$O$_8$ (Y124).\cite{Hussey} A possible explanation for
these discrepancies was suggested based on a systematic study of
Zn-doping effect on the residual thermal conductivity of several
typical cuprate families.\cite{Sun_nonuniversal} By using
exceptionally high quality single-crystal samples and achieving
very small experimental errors, it was found that the ``universal"
picture is not a precise description of the low-$T$ QP transport
properties,\cite{Sun_nonuniversal} whereas the electronic
inhomogeneity was found to be the main reason for the failure of
the classical $d$-wave theory.\cite{Hirschfeld} It should be
mentioned that there is a controversy over the data analysis of
the very-low-$T$ thermal conductivity, that is, whether the
commonly observed $T$-dependence of $\kappa/T$ data below 1 K
weaker than $T^2$ is due to the phonon specular reflection off the
sample surface or simply because the phonon boundary scattering
can only be achieved at temperature lower than $\sim$ 130
mK.\cite{Taillefer, Sutherland_PRB, Sun_logT, Sun_nonuniversal}
The key question is whether the phonons can be scattered by
something else except for the boundary at {\it very low
temperatures}. An indirect but helpful result for this issue was
recently obtained in another novel oxide material
GdBaCo$_2$O$_{5+x}$ (GBCO); the phonon scattering by free spins in
this material was found to be very active in subkelvin
temperatures, which makes the boundary scattering limit not
achievable at temperature as low as 0.3 K.\cite{Sun_GBCO}

The magnetic-field dependence of low-$T$ thermal conductivity is
very informative for the QP transport properties. The pioneering
work in nearly optimally-doped Bi2212 discovered a striking
``plateau" behavior of $\kappa(H)$ isotherms in the temperature
range down to 4 K.\cite{Krishana} It was originally discussed that
the ``plateau" behavior is due to some kind of field-induced phase
transition,\cite{Krishana, Laughlin} which, however, was
questioned by a subsequent experimental observation that the
$\kappa$ value in the ``plateau" depends on the history of
applying the magnetic field.\cite{Aubin_Science} It was then
proposed that this behavior can be well understood by considering
the competition between the QP scattering by vortices, which is
predominant at higher temperature,\cite{Krishana, Franz, Vekhter,
Ando_Bi2212} and the field-induced QP excitations, which is known
as the ``Volovik" effect that is more important at low
temperature.\cite{Volovik, Vekhter} This explanation was supported
by the $\kappa(H)$ measurements at subkelvin
temperatures.\cite{Aubin, Chiao_YBCO} Furthermore, the in-plane
anisotropic heat transport measurements done on Bi2212 single
crystals discovered that the ``plateau" behavior is only present
for heat current along the $b$ axis,\cite{Ando_anisotropy}
probably due to the distortion of $d_{x^2-y^2}$ pairing symmetry.
Subsequently, more efforts were devoted to the studying of
$\kappa(H)$ behaviors at subkelvin temperature for cuprate
superconductors with different doping levels.\cite{Sun_LSCO,
Hawthorn_LSCO, Sun_YBCO, Ando_BSLCO} One interesting finding is
the field-induced suppression of the very-low-$T$ heat
conductivity in the underdoped La$_{2-x}$Sr$_x$CuO$_4$ (LSCO),
whereas the magnetic field naturally enhances the thermal
conductivity of optimally-doped and overdoped LSCO as expected
from the ``Volovik" effect.\cite{Sun_LSCO, Hawthorn_LSCO} The
unusual behavior in the underdoped LSCO can be understood as the
QP localization associated with the magnetic-field-induced spin or
charge order.\cite{Sun_LSCO, Gusynin, Takigawa, Hoffman, Lake,
Khaykovich} Similar $\kappa(H)$ results have been reported for
weakly doped YBCO and BSLCO.\cite{Sun_YBCO, Ando_BSLCO} Note that
in all these previous studies, the field dependence of thermal
conductivity was attributed only to the electron transport and the
contribution of phonon heat conductivity was assumed to be
independent of magnetic field. However, it was recently found that
in an undoped compound Pr$_{1.3}$La$_{0.7}$CuO$_4$ (PLCO), dilute
($\sim$ 1\%) free 1/2-spins (which are presumably created by a
small oxygen nonstoichiometry) can strongly scatter phonons and
produce a pronounced ``dip" behavior in the $\kappa(H)$ isotherms
at low temperatures \cite{Sun_PLCO}. Such an observation implies
that the phonon heat transport could be sensitive to magnetic
field in high-$T_c$ cuprates, which are known to commonly possess
some local magnetic moments.\cite{Moler, Revaz, Wright, Chen,
Brugger, Hien, Schottky} Apparently, correctly understanding the
phonon heat transport property becomes crucial, not only for its
own research interests but also for its role in explaining the
magnetic field dependence of low-$T$ thermal conductivity and the
issue of boundary scattering limit, which are fundamental for
capturing the QP transport properties.

In this work, we study the low-$T$ thermal conductivity of
high-quality Bi2212 single crystals at various dopings to probe
both the QP heat transport and the phonon heat transport
properties. It is worth mentioning that the doping dependence of
the heat transport has been seldom studied for
Bi2212.\cite{Sun_nonuniversal} Besides confirming the decrease of
QP heat transport with decreasing doping by either tuning the
oxygen content or doping Dy, we obtain several novel results: (i)
Dy substitution for Ca stabilizes the crystal lattice and leads to
the improvement of the phonon heat transport; (ii) the low-$T$
phonon peak in $\kappa(T)$ curves is observed in appropriately
Dy-doped samples, which is the first time for Bi2212 system; (iii)
Dy doping introduces some magnetic-field dependence of phonon heat
transport and a ``plateau" phenomenon in $\kappa(H)$ isotherms is
found in the non-superconducting Dy-doped samples.

\section{Experiments}

High-quality pure Bi$_2$Sr$_2$CaCu$_2$O$_{8+\delta}$ and Dy-doped
[Bi$_2$Sr$_2$Ca$_{1-x}$Dy$_x$Cu$_2$O$_{8+\delta}$] single crystals
are grown by the floating zone (FZ) method \cite{Ando_Bi2212,
Ando_anisotropy}. The pure crystals are carefully annealed at 400
-- 800 $^{\circ}$C and in different atmospheres to tune the oxygen
content, {\it e.g.}, the optimal doping with $T_c$ = 95 K is
obtained by annealing at 800 $^{\circ}$C in air. Most of the
Dy-doped samples, if not specially mentioned, are annealed at the
same condition as that of the optimal doping. We label the pure
samples as UD70, OP95 and OD70, etc., for showing their doping
regimes and $T_c$ values, and Dy-doped samples as Dy81, Dy45,
etc., by their $T_c$ values (the Dy doping effectively reduces the
hole concentration and brings the samples into the underdoped
regime). The actual Dy concentration $x$ is determined by the
inductively-coupled plasma atomic-emission spectroscopy (ICP-AES).

The thermal conductivity $\kappa$ is measured along the $a$ axis,
perpendicular to the direction of the incommensurate modulation
structure. The typical size of the sample is $2.5 \times 0.8
\times 0.03$ mm$^3$, where the longest and shortest dimensions are
along the $a$ and $c$ axis, respectively. The measurements are
done by several different processes: using a conventional
steady-state technique by using Chromel-Constantan thermocouple in
a $^4$He cryostat for $\kappa(T)$ data at ``high" temperatures (2
-- 150 K) and for $\kappa(H)$ data above 40 K; using a ``one
heater, two thermometers" technique with Cernox sensors in the
$^4$He cryostat for $\kappa(H)$ data at 5--30 K; using a ``one
heater, two thermometers" technique with RuO$_2$ sensors in a
dilution refrigerator or a $^3$He refrigerator for $\kappa(T)$
data at subkelvin temperatures and for $\kappa(H)$ data below 10
K. All the measurement techniques for $\kappa$ have been described
in our previous publications.\cite{Takeya, Sun_LSCO, Sun_YBCO,
Sun_nonuniversal, Ando_anisotropy, Sun_PLCO} The error and
irreproducibility of thermal conductivity data are typically less
than 10 \%.

Magnetization measurements are carried out using a Quantum Design
SQUID magnetometer.

\section{Results and Discussion}

\subsection{Electrical transport and carrier concentration}

\begin{figure}
\includegraphics[clip,width=8.5cm]{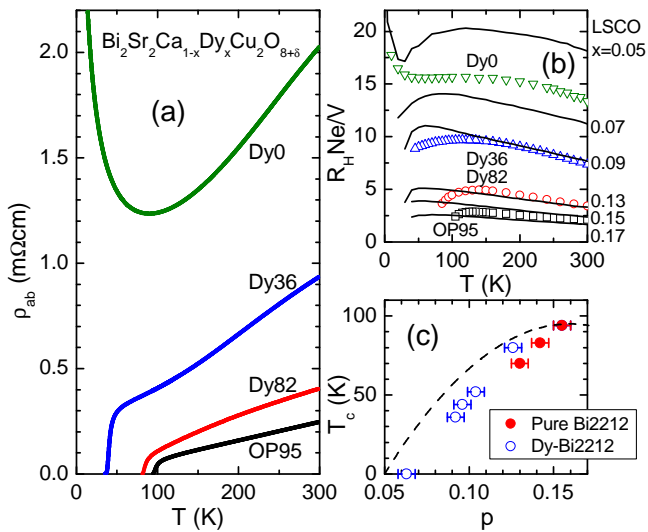}
\caption{(color online) (a) In-plane resistivity of several pure
and Dy-doped Bi2212 single crystals. (b) Renormalized Hall
coefficient of these crystals. LSCO data (solid lines) from Ref.
\onlinecite{Ando_LSCO} are included for comparison. (c) $T_c$ vs
$p$ for several pure and Dy-doped Bi2212 single crystals. Doping
dependence of $T_c$ does not follow the empirical formula
$T_c/T^{max}_c = 1 - 82.6(p-0.16)^2$, shown by the dashed line.}
\end{figure}

Figure 1(a) shows the in-plane resistivity of several Dy-doped
Bi2212 single crystals. The optimally-doped sample (without Dy
doping) shows a ``negative" residual resistivity,
\cite{Ando_Bi2212} suggesting very high cleanliness of our
crystals. Upon doping Dy the superconductivity is gradually
suppressed and the increase of residual resistivity is rather
weak, thus the main reason for the suppression of the
superconductivity is the reduction of hole concentration when
substituting Dy$^{3+}$ ions for Ca$^{2+}$ ions.

It is difficult to determine the carrier concentration in Bi2212
because the hole concentration per Cu, $p$, cannot be chemically
determined like in LSCO. Ando {\it et al.} have shown that the
magnitude of renormalized Hall coefficient $R_HNe/V$, where $N$ is
the number of Cu atoms in a unit cell and $V$ is the volume of the
unit cell, can be used as a tool to identify the $p$
value.\cite{Ando_Hall} Figure 1(b) compares the $R_HNe/V$ of the
Dy-doped Bi2212 samples (Ref. \onlinecite{Abe}) with those of LSCO
(Ref. \onlinecite{Ando_LSCO}), from which the $p$ values can be
easily estimated with rather small uncertainties. Figure 1(c)
shows the obtained $T_c$ vs $p$ relation for several pure and
Dy-doped Bi2212 single crystals.\cite{Abe} Note that the doping
dependence of $T_c$ shows some deviations from the empirical
formula $T_c/T^{max}_c = 1 - 82.6(p-0.16)^2$,\cite{Presland} which
is often used for determining the hole concentration in Bi2212. In
particular, the boundary between non-superconducting and
superconducting phases turns out to be located at $p \approx$
0.07.

\subsection{Temperature dependence of thermal conductivity at ``high" temperature}

\begin{figure}
\includegraphics[clip,width=5.5cm]{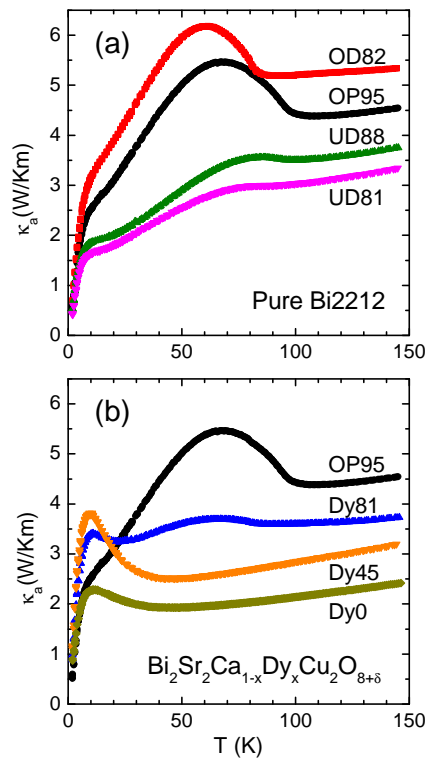}
\caption{(color online) Temperature dependence of thermal
conductivity of pure and Dy-doped Bi2212 single crystals. The Dy
contents $x$ for Dy81, Dy45 and Dy0 samples are 0.17, 0.30 and
0.53, respectively.}
\end{figure}

Figure 2 shows the thermal conductivities of several pure and
Dy-doped Bi2212 single crystals from 2 to 150 K. The
optimally-doped sample shows a well-known broad ``QP peak" below
$T_c$ as well as a ``knee" feature around 10 K, whose origin has
never been clarified. The size of QP peak (the enhancement from
the minimum near $T_c$ to the peak is 1.1 W/Km or 25 \%) is among
the largest in Bi2212 single crystals that have ever been
studied,\cite{Krishana, Ando_Bi2212, Ando_anisotropy, Bi2212}
confirming the high quality of the crystals. In pure samples,
either increasing or decreasing the hole concentration results in
the weakening of the QP peak while the magnitude of total thermal
conductivity monotonously decreases with decreasing the hole
concentration. Furthermore, the peak suppression is much more
significant in the underdoped samples than that in the overdoped
samples, which is possibly related to the nanoscale inhomogeneity
found by STM experiments in the superconducting state of this
compound.\cite{Howald, Lang, McElroy} This behavior is very
similar to that in the Zn-doped Bi2212,\cite{Ando_Bi2212} in which
local suppression of the superconductivity leading to a patchy
superconducting state is playing a role.\cite{Pan} Upon doping Dy,
the QP peak is also strongly suppressed but a new peak at 10 K
emerges. Interestingly, in the Dy81 ($x$ = 0.17) sample, both the
QP peak and the 10 K peak show up, which means that the 10 K peak
is not directly related to the QP heat transport. This low-$T$
peak is the strongest in the Dy45 ($x$ = 0.30) sample and survives
in the non-superconducting sample (Dy0, $x$ = 0.53), which clearly
indicates that it originates from the phonon heat transport.

In the undoped compound of other cuprates, like La$_2$CuO$_4$,
YBa$_2$Cu$_3$O$_6$, Nd$_2$CuO$_4$ and Pr$_{1.3}$La$_{0.7}$CuO$_4$,
the phonon heat transport behaves similarly to that of common
insulators and shows a large phonon peak around 20 K, whose
magnitude varies between $\sim$ 20 and $\sim$ 80 W/Km.\cite{Cohn,
Sun_LCO, Jin, Sun_PLCO, Berggold} However, it is known that Bi2212
system has much dirtier phonon heat transport than other
cuprates,\cite{Bi2212} probably due to the strong disorder of the
crystal lattice caused by the excess oxygen and the incommensurate
superlattice along the $b$ axis. The incommensurate modulation had
been discussed to be related to the extra oxygen locating in BiO
double layers and the lattice mismatch between the BiO block and
the CuO$_2$ layer.\cite{Kambe} In an earlier work, the thermal
conductivity of Bi2212 parent material was studied in
Bi$_2$Sr$_2$YCu$_2$O$_8$ and it was found that the phonon heat
transport is too weak to give rise to the phonon peak (only a
hump-like feature observed at 20 -- 30 K with the magnitude of
$\kappa$ smaller than 2 W/Km).\cite{Bi2212} To our knowledge, the
present result shows the phonon peak of Bi2212 for the first time.
It is not surprising that the phonon peak is located at much lower
temperature than in other cuprates if the defect scattering is
much stronger in Bi2212. The present result also indicates that a
moderate Dy doping can enhance the phonon conductivity rather
strongly despite that the atomic substitution always introduces
additional impurity/defect scattering on phonons. Apparently, the
crystal structure is somewhat stabilized by doping an appropriate
amount of Dy. A possible scenario is that substituting Ca$^{2+}$
ions with Dy$^{3+}$ increases the oxygen content and results in
the expansion of the BiO bolck, which is helpful for relaxing the
lattice mismatch between the BiO block and CuO$_2$
plane.\cite{Kambe} Note that the low-$T$ peak observed in the
Dy-doped samples clarifies that the ``knee" feature of $\kappa(T)$
in superconducting Bi2212 single crystals is due to the
competition between the decrease of QP heat transport and the
increase of phonon heat transport upon lowering temperature across
10 K. Furthermore, the phonon heat transport in pure Bi2212 is
obviously very weak, and its magnitude at 10 K (where the phonon
conductivity peaks) is smaller than or at most comparable to that
in the non-superconducting Dy-Bi2212.

\begin{figure}
\includegraphics[clip,width=5.5cm]{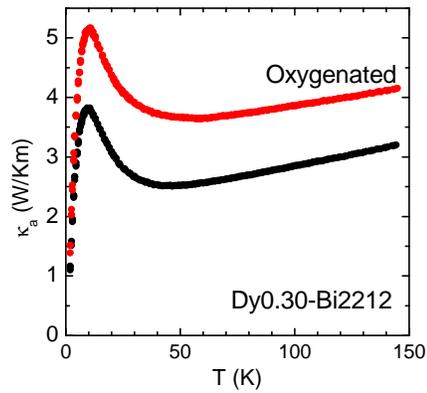}
\caption{(color online) Thermal conductivity of Dy-doped Bi2212
($x$ = 0.30) single crystal (annealed at 800 $^{\circ}$C in air)
and the same sample after oxygenating heat treatment (annealed at
500 $^{\circ}$C in air).}
\end{figure}

Similar to the pure Bi2212 crystals, the heat transport of
Dy-Bi2212 is also sensitive to the oxygen content. After
appropriate annealing to increase the oxygen content, for example,
the $x$ = 0.30 crystal (Dy45 in Fig. 2) presents a
parallel-shift-like enhancement of the thermal conductivity at
first glance, as shown in Fig. 3. However, we should note that the
thermal conductivity enhancement above $T_c$ is mainly caused by
the increase of the electron heat transport (which can be roughly
estimated from the normal-state resistivity data using the
Wiedemann-Franz law), while the enhancement of the 10K-peak should
be mainly due to the improvement of phonon transport since the QP
(or electron) heat transport in this temperature region is
expected to be strongly damped in highly Dy-doped samples. It is
likely that the oxygenating process reduces the number of oxygen
vacancies in the crystal, which scatter phonons. The magnitude of
the phonon peak at 10 K increases from 3.8 to 5.2 W/Km, but both
curves do not show the QP peak although the superconducting
temperature increases from 45 to 55 K.

\subsection{Temperature dependence of thermal conductivity at subkelvin temperature}

To further investigate the doping dependence of QP heat transport,
the thermal conductivities of Bi2212 and Dy-doped Bi2212 single
crystals are measured at very low temperatures down to 70 mK. Some
of the data have been reported in a previous publication (see
Figs. 3 and 4 in Ref. \onlinecite{Sun_nonuniversal}). It has been
shown that in the lowest-$T$ regime where phonons are in the
boundary scattering limit, the QP term and the phonon term of
thermal conductivity can be separated by fitting data (below
$\sim$ 130 mK) to
\begin{equation}
\frac{\kappa}{T} = \frac{\kappa_0}{T} + bT^2, \label{T^2}
\end{equation}
where the residual term $\kappa_0/T$ is the QP contribution and
$bT^2$ is the phonon contribution.\cite{Taillefer, Takeya,
Sun_YBCO, Sun_nonuniversal} Note that the residual thermal
conductivity of a $d$-wave superconductor was discussed to present
a ``universal" behavior, that is, it is independent of the QP
scattering rate if the impurity bandwidth satisfies $k_BT \ll
\gamma \ll \Delta_0$ ($\Delta_0$ is the maximum gap).\cite{Graf,
Durst} In this picture, the QP heat conductivity ($\kappa_e/T$) of
cuprate superconductor decreases with temperature and finally
arrives at a certain value, which depends only on the doping
level, at low enough temperature. However, our Bi2212 crystals
clearly show much larger residual thermal conductivity compared to
the samples used in earlier works.\cite{Chiao, Behnia} Judging
from the methods of growing single crystals and the normal-state
resistivity data,\cite{Chiao, Behnia} it is obvious that the
difference in $\kappa_0/T$ is strongly related to the cleanliness
of the crystals, which actually challenges the validity of the
``universal" behavior. Indeed, this issue was recently elucidated
by a critical examination of the ``universal" behavior in a
well-controlled experiment;\cite{Sun_nonuniversal} it was found
that the disorder or inhomogeneity of cuprates, which are
especially strong in Bi2212, make the ``universal" behavior to be
only a rough approximation. Nonetheless, the residual thermal
conductivity universally shows a monotonous doping dependence in
all cuprates.

Recently, a controversy arose for the data analysis of the
very-low-$T$ thermal conductivity of high-$T_c$ cuprates; namely,
whether the commonly observed $T$-dependence of $\kappa/T$ data
below 1 K weaker than $T^2$ is due to the specular reflection of
phonons off the sample surface or simply because the phonon
boundary scattering can only be achieved at temperatures lower
than $\sim$ 130 mK.\cite{Taillefer, Sutherland_PRB, Sun_logT,
Sun_nonuniversal} In this regard, although Taillefer's group was
the first to propose the analysis of Eq. (\ref{T^2}) and
successfully described the very-low-$T$ data of
YBCO,\cite{Taillefer} they have recently shifted to conjecture the
existence of specular reflections in a wide temperature range and
analyzed their data of YBCO and LSCO by fitting to
\begin{equation}
\frac{\kappa}{T} = \frac{\kappa_0}{T} + aT^\gamma, \label{T^gamma}
\end{equation}
where the second term with $\gamma <$ 2 is purported to be the
phonon conductivity in the specular scattering
limit.\cite{Sutherland_PRB} However, it was recently pointed
out\cite{Ando_Condmat} that one can estimate the phonon mean-free
path $\ell_p$ in the temperature range of interest, which shows
that that $\ell_p$ is too short for the specular-reflection
scenario to be valid. Furthermore, the analysis involving Eq. (3)
analysis is actually based on another assumption that the QP
thermal conductivity is independent of temperature up to at least
$\sim$ 1 K, which is inconsistent with an experimental finding
from the same group\cite{Hill} that in YBCO the QP thermal
conductivity at finite temperature increases quickly with
temperature, following a $T^3$ dependence.\cite{note} In this
sense, the ``good" fitting of Eq. (\ref{T^gamma}) to some
experimental data are not physically meaningful.

\begin{figure}
\includegraphics[clip,width=5.5cm]{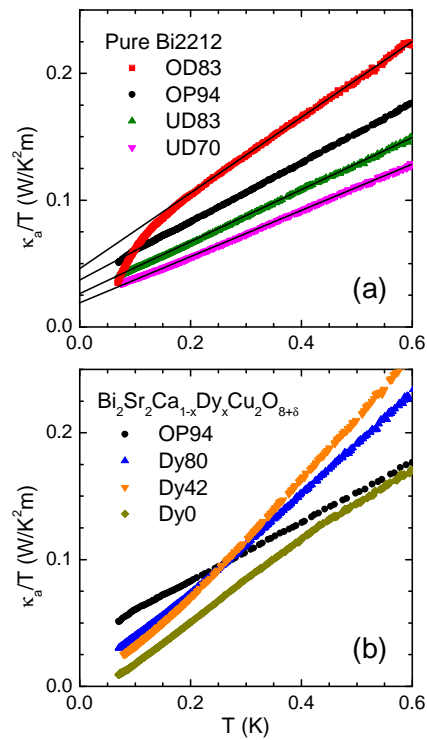}
\caption{(color online) Low-$T$ thermal conductivity of pure and
several Dy-doped Bi2212 single crystals measured down to 70 mK,
plotted as $\kappa/T$ vs $T$. The Dy contents $x$ for Dy80, Dy42
and Dy0 samples are 0.20, 0.34 and 0.51, respectively. Strong
downturn of the low-$T$ data of OD83 sample is caused by the
electron-phonon decoupling.}
\end{figure}

If one tries to analyze the data of Bi2212 using Eq.
(\ref{T^gamma}), it can be found that the low-$T$ thermal
conductivity actually follows
\begin{equation}
\frac{\kappa}{T} = \frac{\kappa_0}{T} + AT, \label{T}
\end{equation}
below 0.6 K down to nearly 100 mK, as shown in Fig. 4; here the
power of $T$ in the second term is clearly too small to be
attributed to the phonon specular reflection of
phonons.\cite{Sutherland_PRB} Note that the same temperature
dependence of $\kappa$ has recently been observed in
Tl$_2$Ba$_2$CuO$_{6+\delta}$ (Tl2201) system, whose doping levels
are located in the overdoped regime.\cite{Hawthorn_Tl} Hawthorn
{\it et al.} attributed the phonon conductivity of $AT$ to the
dominant electron-phonon scattering with high carrier
concentration of Tl2201,\cite{Hawthorn_Tl} like the case in a
metal, which actually means that {\it the phonon boundary
scattering limit can never be established in this system}. Our
present results, on the other hand, are observed mostly in the
{\it underdoped or optimally doped} regions of Bi2212, indicating
that the $AT$ term may not be the simple phonon contribution.
Considering both the much larger $\kappa_0/T$ term of Bi2212
compared to other cuprates and the much weaker phonon transport
seen in the ``high"-$T$ region (Fig. 2), it is reasonable to
conclude that the $T^2$ dependence of the low-$T$ thermal
conductivity (above $\sim$ 100 mK) of Bi2212 is related to an
amorphous-like phonon transport \cite{Berman} together with a
$T^2$-dependent QP heat conductivity. This is essentially
consistent with a previous finding in Bi2212 by Nakamae {\it et
al.}\cite{Behnia}

To determine the residual term $\kappa_0/T$ at zero temperature by
employing some extrapolation of the data at finite temperature,
one should always keep in mind that the formula used for the
extrapolation must have a physical reasoning, which assures that
the same functional form holds throughout the range of the
extrapolation (to 0 K). Therefore, although Eq. (\ref{T}) can fit
the Bi2212 data pretty well in a broad temperature range, this
fitting {\it cannot} provide the precise determination of
$\kappa_0/T$, because the $T^2$-dependent QP heat conductivity
does not hold at very low temperature ($\sim$ 100 mK or lower)
when the temperature scale becomes smaller than the impurity
bandwidth.\cite{Graf, Behnia} For this reason, it is still the
most reliable way to obtain $\kappa_0/T$ by fitting the lowest-$T$
data to Eq. (\ref{T^2}), as we did in Ref.
\onlinecite{Sun_nonuniversal}.

In passing, it should be pointed out that the very-low-$T$ data of
overdoped Bi2212 show a significant downturn below 0.2 K, which is
known to be due to the electron-phonon decoupling at very low
temperatures\cite{Smith} that is often significant in the
overdoped cuprates.\cite{Kim} This phenomenon further demonstrates
that the $T^2$ behavior of low-$T$ thermal conductivity in this
system does not have the same origin as that in Tl2201, which was
discussed to be the strong electron-phonon
scattering.\cite{Hawthorn_Tl}

\subsection{Magnetic-field dependence of thermal conductivity}

\begin{figure}
\includegraphics[clip,width=8.5cm]{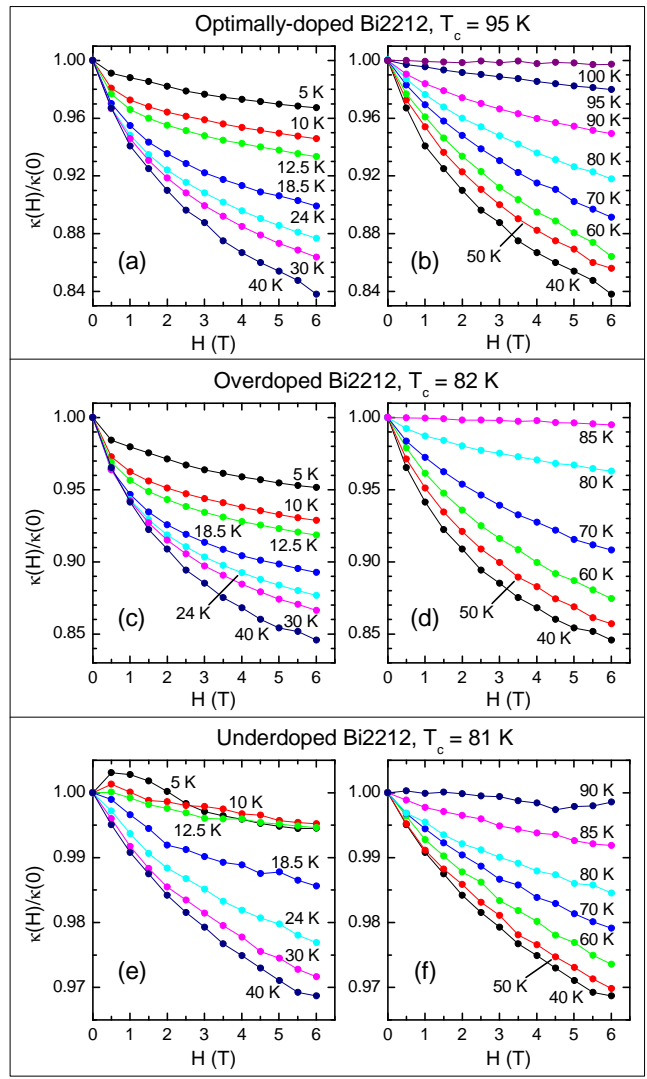}
\caption{(color online) Magnetic-field dependences of the low-$T$
thermal conductivity of the pure Bi2212 single crystals, whose
doping levels cover from underdoped to overdoped regions. The
field direction is along the $c$ axis.}
\end{figure}

The magnetic-field dependence of both pure and Dy-doped Bi2212
crystals are studied in detail for $H \parallel c$. Figures 5(a)
and 5(b) show the results of OP95 sample at 5 -- 100 K, which are
essentially consistent with previous studies.\cite{Krishana,
Ando_Bi2212, Ando_anisotropy} Note that the famous ``plateau"
feature is absent in these $\kappa(H)$ isotherms since the thermal
conductivity is always measured along the $a$ axis in the present
work. As shown in Fig. 6, the field-dependence of $\kappa$ can be
well described by the simple relation
\begin{equation}
\kappa(H,T) = \frac{k_e(T)}{1 + p(T)H} + \kappa_{ph}(T),
\label{kH}
\end{equation}
where $\kappa_e$ and $\kappa_{ph}$ are the electronic thermal
conductivity and the phononic thermal conductivity, respectively;
here the field dependence is attributed to the quasiparticle
scattering off vortices.\cite{Ando_Bi2212} Apparently, the
amplitude of the field dependence of $\kappa$ is dependent on the
contribution of QPs to the total thermal conductivity. One should
notice that the magnetic field dependence of QP thermal
conductivity in $d$-wave cuprates is determined by the competition
between the decrease of heat conductivity due to the vortex
scattering and the increase of heat conductivity due to the
increase of the nodal QP excitations, the so-called ``Volovik"
effect.\cite{Franz, Vekhter, Ando_Bi2212, Volovik, Aubin,
Chiao_YBCO} The former effect is dominant at high temperature
whereas the latter one is more important at low temperature. Thus,
the weakliness of the field dependence of $\kappa$ at low
temperatures could be due either to the decrease of QP portion of
heat conductivity or to the ``Volovik" effect, or both.

\begin{figure}
\includegraphics[clip,width=8.5cm]{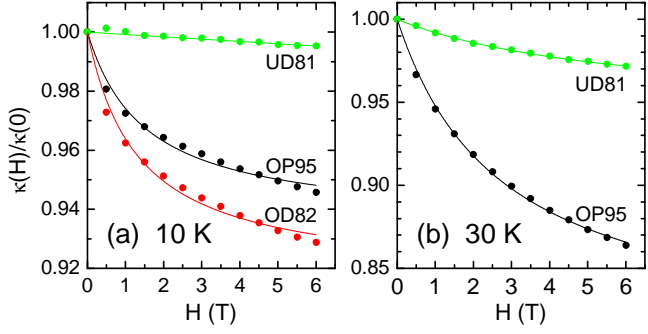}
\caption{(color online) Representative $\kappa(H)$ isotherms of
pure Bi2212 single crystals with different doping level and the
corresponding theoretical fitting curves using Eq. (\ref{kH}).}
\end{figure}

As shown in Figs. 5(c--f), the field dependence of $\kappa$ is
also rather strong in the overdoped sample, almost the same as
that in the optimally-doped sample, whereas it is weakened by
several times in the underdoped sample. Nevertheless, the
$\kappa(H)$ data in these pure Bi2212 crystals at all temperatures
studied are well described by Eq. (\ref{kH}) (see Fig. 6 for
several representative data), which means the QP scattering
mechanism does not change upon tuning the oxygen content.
Apparently, the weakening of $\kappa(H)$ in underdoped sample is
due to the suppression of the QP contribution to the heat
conductivity, consistent with the conclusion from the
temperature-dependence data.

Note that the above discussion is based on an assumption that the
phonon thermal conductivity is independent of magnetic field,
which however has recently been found to be not always valid for
high-$T_c$ cuprates. In a parent compound of high-$T_c$ cuprates,
PLCO, in which the phononic heat transport is very clean, diluted
free spins associated with a small oxygen nonstoichiometry
introduces strong magnetic-field dependence of phonon thermal
conductivity through the phonon scattering off free
spins.\cite{Sun_PLCO} Since the localized free spins are commonly
found in superconducting cuprates, as is evidenced by the Schottky
anomaly in the specific heat,\cite{Moler, Revaz, Wright, Chen,
Brugger, Hien, Schottky} some fraction of the magnetic-field
dependence of $\kappa$ is likely due to the spin-phonon
scattering. In the case of pure Bi2212, however, the effect of
spin scattering on the phonon heat transport would be unimportant
because the phonons are strongly scattered by the lattice disorder
or defects.

\begin{figure}
\includegraphics[clip,width=8.5cm]{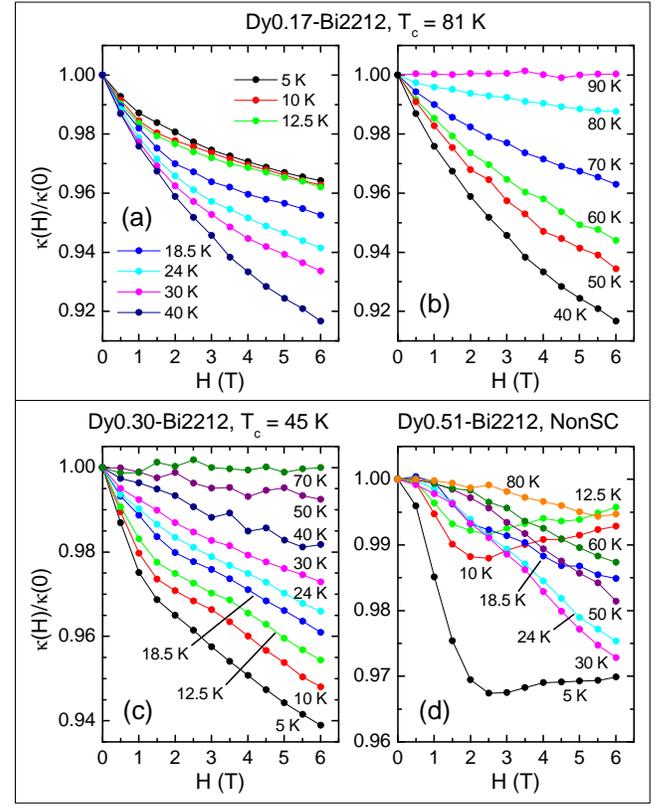}
\caption{(color online) Magnetic-field dependences of the low-$T$
thermal conductivity of the Dy-Bi2212 ($x$ = 0.17, 0.30 and 0.51)
single crystals, which are superconducting at 81 K, 45 K and
non-superconducting, respectively.}
\end{figure}

The magnetic-field dependences of the thermal conductivity of
Dy-doped samples with $x$ = 0.17, 0.30 and 0.51 are shown in Fig.
7. Similar to the case of Dy-free crystals, the field dependence
of $\kappa$ is gradually weakened with decreasing carrier
concentration, which should be also due to the reduction of the
electronic contribution to the heat conductivity. An interesting
point is that the $\kappa$ in Dy-Bi2212 samples show more
complicated magnetic-field dependences. For $x$ = 0.17, the
magnetic-field dependence of $\kappa$ can still be well described
by Eq. (\ref{kH}). On the other hand, for $x$ = 0.30, an
``S"-shaped feature at 1 -- 3 T shows up in the $\kappa(H)$
isotherms below 30 K; the fitting to these data using Eq.
(\ref{kH}), as shown in Fig. 8, suggests that there is an
additional dip-like change of $\kappa$ superimposed on the
background of usual field dependence described by Eq. (\ref{kH})
and the position of the dip slightly moves to lower field with
decreasing temperature. For the $x$ = 0.51 sample, in which the
overall field dependence is much weaker, a step-like decrease of
$\kappa$ clearly occurs around 2 T at low temperatures.

\begin{figure}
\includegraphics[clip,width=8.5cm]{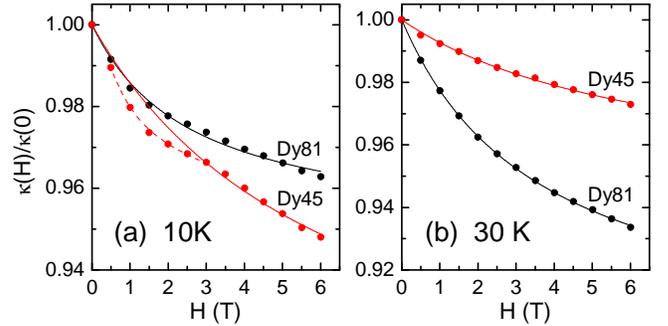}
\caption{(color online) Representative $\kappa(H)$ isotherms of
Dy-doped Bi2212 single crystals with $x$ = 0.17 and 0.30, and the
corresponding theoretical fitting curves (solid lines) using Eq.
(\ref{kH}). The dashed line indicates the deviation of
experimental data from the theoretical curve.}
\end{figure}

To probe whether the peculiar $\kappa(H)$ behavior for higher Dy
dopings is caused by the oxygen defects, the field dependences of
$\kappa$ are also measured for the $x$ = 0.30 sample after
re-annealed at 500 $^{\circ}$C in air, which increases the oxygen
content and increases the $T_c$ from 45 K to 55 K (the temperature
dependence of $\kappa$ is shown in Fig. 3). It can be seen from
Figs. 7(c) and 9 that the profiles of $\kappa(H)$ do not change so
much with increasing oxygen content except that the dip-like
feature becomes more pronounced at low temperatures. The position
of the dip is still located at 1 -- 3 T. This indicates that the
oxygen defects are not directly related to the physical origin of
this peculiar feature.

\begin{figure}
\includegraphics[clip,width=5cm]{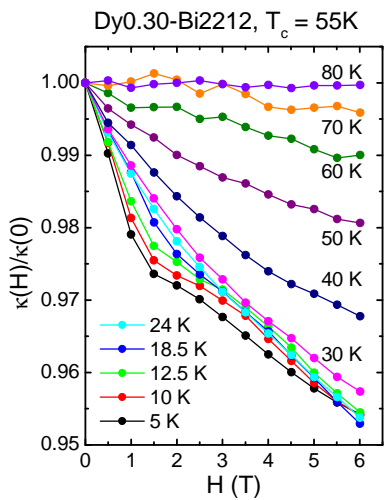}
\caption{(color online) Magnetic-field dependences of the low-$T$
thermal conductivity of $x$ = 0.30 sample after re-annealed at 500
$^{\circ}$C in air.}
\end{figure}

Since both the quasiparticle transport and the oxygen defects are
not likely to be producing the additional contribution to the
field-dependence of $\kappa$, it is natural to conclude that the
Dy$^{3+}$ ions are playing the key role. Apparently, the $x$ =
0.30 sample does not have any long-range order of Cu spins or Dy
spins, so the Dy$^{3+}$ ions are most likely affecting the heat
transport as free spins. A common magnetic scattering on phonons
is caused by free spins or paramagnetic moments through
spin-phonon interaction.\cite{Berman} In this case the energy
splitting of the spin states that induces resonant scattering on
phonons is increased by the Zeeman effect. The phonon scattering
off these spins is most effective in suppressing the heat
transport when the energy splitting is equal to $\sim$ 3.8 $k_BT$,
where the spectrum of phonon heat conductivity
peaks;\cite{Berman,Sun_GBCO} therefore, the spin-phonon scattering
usually generates a dip-like feature in $\kappa(H)$ at this energy
and the dip position shifts to higher fields with increasing
temperature.\cite{Berman} As we have mentioned, the effect of the
magnetic phonon scattering on the heat transport has been known
for a long time and has recently been found to affect the
field-dependence of $\kappa$ in cuprates.\cite{Sun_PLCO} While the
magnetic scattering on phonons seems unimportant in pure Bi2212
single crystals, in which the phonon heat transport is strongly
damped by defects and lattice disorders, it becomes much more
pronounced upon doping Dy that not only brings free spins but also
enhances the phonon heat transport.

As shown in Fig. 7(d), a step-like $\kappa(H)$ isotherms is
observed at low temperatures in the $x$ = 0.51 sample. Is this
also due to the phonon scattering by free spins? In this scenario,
if the energy splitting is larger than 3.8 $k_BT$, the magnetic
field further increases the energy splitting and simply weakens
the scattering and produces a step-like {\it increase} of thermal
conductivity.\cite{Sun_GBCO} Apparently, the step-like {\it
decrease} of $\kappa(H)$ in Dy-Bi2212 is out of the expectation
from the common spin-phonon scattering mechanism and suggests some
other physical mechanism. Since the $x$ = 0.51 sample is
non-superconducting with an insulating resistivity behavior at low
temperatures, the step-like $\kappa(H)$ must also be irrelevant to
the electron transport.

\subsection{``Plateau" of low-temperature $\kappa(H)$ isotherms
in Dy-Bi2212}

\begin{figure}
\includegraphics[clip,width=8.5cm]{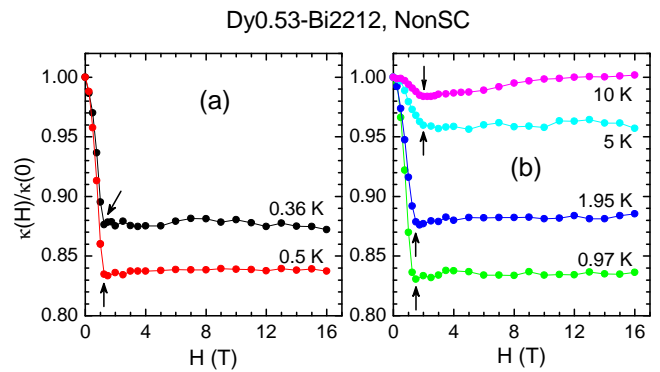}
\caption{(color online) Magnetic-field dependences of the low-$T$
thermal conductivity of the Dy-Bi2212 ($x$ = 0.53) single crystal
down to 0.36 K. The magnetic field is applied along the $c$ axis.}
\end{figure}

In Fig. 7(d), the $\kappa(H)$ isotherm of $x$ = 0.51 sample at 5 K
shows a step-like decrease at low field and $\kappa$ becomes
nearly independent of magnetic field above 2 T, a ``plateau"-like
behavior. We note that it is not clear whether this ``plateau"
behavior is the same as that observed in pure Bi2212 along the $b$
axis,\cite{Ando_anisotropy} because the latter is believed to have
its origin in the QP transport, while in the present case the
sample is an insulator. To investigate this intriguing behavior in
Dy-doped Bi2212, the magnetic-field dependence of $\kappa$ are
studied in more detail at very low temperature down to 0.36 K and
in high magnetic field up to 16 T (along the $c$ axis). As shown
in Fig. 10, the ``plateau"-like behavior is well reproduced in
another sample of $x$ = 0.53 at low temperatures. With lowering
temperature, the step-like suppression of $\kappa$ becomes more
significant and the strongest ($\sim$ 16\%) suppression is
observed between 0.97 and 0.5 K. At even lower temperature of 0.36
K, the field-induced suppression of $\kappa$ is getting weaker.
The characteristic field, where the suppression completes or the
``plateau" starts, is weakly dependent on the temperature,
increases slightly from $\sim$ 1 T at 0.36 K to $\sim$ 2 T at 5 K.
Somewhat intriguingly, this trend is the same as that found for
the plateau in the superconducting Bi2212 samples.\cite{Krishana}

\begin{figure}
\includegraphics[clip,width=8.5cm]{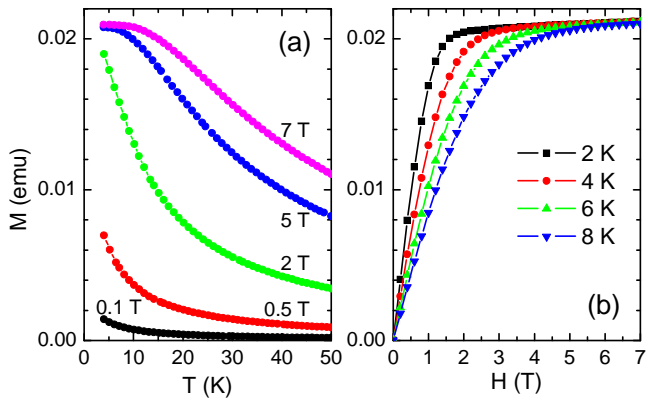}
\caption{(color online) Temperature and field dependences of
magnetization of a non-superconducting Dy-Bi2212 single crystal
with $x$ = 0.53.}
\end{figure}

Figure 11 shows the temperature and field dependences of
magnetization for an $x$ = 0.53 sample. It is clear that the
magnetization data show rather conventional spin polarization of
paramagnetic spins, which must be spins of Dy$^{3+}$ ions. The
characteristic field for spin polarization is about 2 T at 2 K and
increases quickly with increasing temperature; this is rather
different from the characteristic field of $\kappa(H)$ where the
``plateau" appears. In this sense, the $\kappa(H)$ behavior does
not appear to have direct relation with the Dy$^{3+}$ free spins
and their polarization, which is different from the case of the
$x$ = 0.30 sample.

It is known that in rare-earth- or Y-doped Bi2212,\cite{Gao} the
antiferromagnetic (AF) order of Cu spins starts to be established
when the superconductivity is completely suppressed and charge
carriers are localized upon increasing the concentration of
rare-earth ions. Therefore, one possible picture of the plateau
behavior would be the magnon heat transport being affected by
applied magnetic field: In a long-range-ordered spin system, the
low-energy magnon excitations can act as heat carriers, but when
the magnetic field drives the magnon states to sufficiently high
energies, these magnons are no longer thermally excited and drop
from heat conduction.\cite{Walton} This causes a plateau-like
$\kappa(H)$ behavior in magnetic systems, as has been observed in
the yttrium iron garnet system.\cite{Walton} However, one should
keep in mind that the magnons are possibly involved only when the
antiferromagnetic state has un-gapped (or very weakly gapped)
acoustic magnon branches, which needs to be confirmed by neutron
measurements. More importantly, the very abrupt nature of the
onset of the plateau observed at very low temperature [Fig.
10(a)], together with the rather weak temperature dependence of
the magnetic field from which the plateau starts, seems to be
difficult to be understood within this picture. Also, it is even
not clear whether the ``plateau" is only observable in
non-superconducting Dy-doped samples. In other word, although the
``plateau" feature is not present in the low-$T$ $\kappa(H)$
curves for the $a$-axis Bi2212 samples in strong fields up to 16
T,\cite{Ando_anisotropy} the present data could not clarify
whether it is also absent in high magnetic fields for
superconducting (and not AF ordered) Dy-Bi2212 with lower
concentration of $x$. Detailed future study of this plateau
phenomenon is clearly called for.

In passing, we note that the ``plateau" behavior gradually weakens
with increasing temperature, which is possibly due to the
disappearance of the antiferromagnetic order. However, at even
higher temperatures, the field dependence becomes again stronger
above 18.5 K, as is shown in Fig. 7(d). The strongest
$H$-dependence is observed at 30 K, where the magnetic field of 6
T induces $\sim$ 3 \% suppression of $\kappa$ without any
signature of saturation. This magnetic-field dependence of
$\kappa$ at high temperature is actually difficult to understand
because if phonon transport is not affected by magnetic field, the
only way to bring the field dependence of $\kappa$ is through the
electron channel. In this regard, the magnetoresistance
measurements have shown that the in-plane resistivity changes by
less than 0.1 \% in magnetic field ($\parallel c$) up to 14
T.\cite{Abe} Such small change of resistivity cannot bring several
percent change of thermal conductivity unless the Wiedemann-Franz
law is strongly violated in this case.

\section{Summary}

Both the quasiparticle and phonon heat transport of
Bi$_2$Sr$_2$CaCu$_2$O$_{8+\delta}$ are carefully studied for a
broad doping range by using pure Bi2212 single crystals with tuned
oxygen contents and
Bi$_2$Sr$_2$Ca$_{1-x}$Dy$_x$Cu$_2$O$_{8+\delta}$ single crystals
with different Dy contents. It is found that the QP transport
decreases with lowering doping by either decreasing oxygen content
or increasing Dy content, whereas the phonon conductivity is
enhanced in moderately Dy-doped samples where a phonon peak is
observed at 10 K, which means that Dy$^{3+}$ ions not only act as
impurities or point defects but also stabilize the crystal
lattice. The magnetic-field dependence of $\kappa$ in pure Bi2212
at temperature above 5 K is mainly due to the QP scattering off
vortices. In contrast, the Dy-doped samples (with $x$ = 0.30)
present an additional ``dip"-like term of $\kappa(H)$ at low
field, which is discussed to be related to the phonon scattering
by free spins of Dy$^{3+}$ ions. In non-superconducting Dy-Bi2212
samples with $x \simeq$ 0.50, a ``plateau"-like $\kappa(H)$
isotherms with characteristic field at 1 -- 2 T is observed. We
discuss that this ``plateau" is possibly related to the magnon
transport.

\begin{acknowledgments}
This work was supported by the National Natural Science Foundation
of China (10774137 and 50421201), the National Basic Research
Program of China (2006CB922005), and KAKENHI 16340112 and
19674002.
\end{acknowledgments}


\begin{thebibliography}{}

\bibitem{Hussey_Review}
For a review, see N. E. Hussey, Adv. Phys. {\bf 51}, 1685 (2002).

\bibitem{Graf}
M. J. Graf, S-K. Yip, J. A. Sauls, and D. Rainer, Phys. Rev. B
{\bf 53}, 15147 (1996).

\bibitem{Durst}
A. C. Durst and P. A. Lee, Phys. Rev. B {\bf 62}, 1270 (2000).

\bibitem{Taillefer}
L. Taillefer, B. Lussier, R. Gagnon, K. Behnia, and H. Aubin,
Phys. Rev. Lett. {\bf 79}, 483 (1997).

\bibitem{Behnia}
S. Nakamae, K. Behnia, L. Balicas, F. Rullier-Albenque, H. Berger,
and T. Tamegai, Phys. Rev. B {\bf 63}, 184509 (2001).

\bibitem{Chiao}
M. Chiao, R. W. Hill, C. Lupien, L. Taillefer, P. Lambert, R.
Gagnon, and P. Fournier, Phys. Rev. B {\bf 62}, 3554 (2000).

\bibitem{Sutherland_PRB}
M. Sutherland, D. G. Hawthorn, R. W. Hill, F. Ronning, S.
Wakimoto, H. Zhang, C. Proust, E. Boaknin, C. Lupien, L.
Taillefer, R. Liang, D. A. Bonn, W. N. Hardy, R. Gagnon, N. E.
Hussey, T. Kimura, M. Nohara, and H. Takagi, Phys. Rev. B {\bf
67}, 174520 (2003).

\bibitem{Hawthorn_Tl}
D. G. Hawthorn, S. Y. Li, M. Sutherland, E. Boaknin, R. W. Hill,
C. Proust, F. Ronning, M. A. Tanatar, J. Paglione, L. Taillefer,
D. Peets, R. Liang, D. A. Bonn, W. N. Hardy, and N. N. Kolesnikov,
Phys. Rev. B {\bf 75}, 104518 (2007).

\bibitem{Takeya}
J. Takeya, Y. Ando, S. Komiya, and X. F. Sun, Phys. Rev. Lett.
{\bf 88}, 077001 (2002).

\bibitem{Sun_YBCO}
X. F. Sun, K. Segawa, and Y. Ando, Phys. Rev. Lett. {\bf 93},
107001 (2004).

\bibitem{Sun_logT}
X. F. Sun, K. Segawa, and Y. Ando, Phys. Rev. B {\bf 72},
100502(R) (2005).

\bibitem{Ando_BSLCO}
Y. Ando, S. Ono, X. F. Sun, J. Takeya, F. F. Balakirev, J. B.
Betts, and G. S. Boebinger, Phys. Rev. Lett. {\bf 92}, 247004
(2004).

\bibitem{Hussey}
N. E. Hussey, S. Nakamae, K. Behnia, H. Takagi, C. Urano, S.
Adachi, and S. Tajima, Phys. Rev. Lett. {\bf 85}, 4140 (2000).

\bibitem{Sun_nonuniversal}
X. F. Sun, S. Ono, Y. Abe, S. Komiya, K. Segawa, and Y. Ando,
Phys. Rev. Lett. {\bf 96}, 017008 (2006).

\bibitem{Hirschfeld}
W. A. Atkinson and P. J. Hirschfeld, Phys. Rev. Lett. {\bf 88},
187003 (2002).

\bibitem{Sun_GBCO}
X. F. Sun, A. A. Taskin, X. Zhao, A. N. Lavrov, and Y. Ando, Phys.
Rev. B (in press).

\bibitem{Krishana}
K. Krishana, N. P. Ong, Q. Li, G. D. Gu, and N. Koshizuka, Science
{\bf 277}, 83 (1997).

\bibitem{Laughlin}
R. B. Laughlin, Phys. Rev. Lett. {\bf 80}, 5188 (1998).

\bibitem{Aubin_Science}
H. Aubin, K. Behnia, S. Ooi, and T. Tamegai, Science {\bf 280}, 9a
(1998).

\bibitem{Franz}
M. Franz, Phys. Rev. Lett. {\bf 82}, 1760 (1999).

\bibitem{Vekhter}
I. Vekhter and A. Houghton, Phys. Rev. Lett. {\bf 83}, 4626
(1999).

\bibitem{Ando_Bi2212}
Y. Ando, J. Takeya, Y. Abe, K. Nakamura, and A. Kapitulnik, Phys.
Rev. B {\bf 62}, 626 (2000).

\bibitem{Volovik}
G. E. Volovik, JETP Lett. {\bf 58}, 469 (1993).

\bibitem{Aubin}
H. Aubin, K. Behnia, S. Ooi, and T. Tamegai, Phys. Rev. Lett. {\bf
82}, 624 (1999).

\bibitem{Chiao_YBCO}
M. Chiao, R. W. Hill, C. Lupien, B. Popic, R. Gagnon, and L.
Taillefer, Phys. Rev. Lett. {\bf 82}, 2943 (1999).

\bibitem{Ando_anisotropy}
Y. Ando, J. Takeya, Y. Abe, X. F. Sun, and A. N. Lavrov, Phys.
Rev. Lett. {\bf 88}, 147004 (2002).

\bibitem{Sun_LSCO}
X. F. Sun, S. Komiya, J. Takeya, and Y. Ando, Phys. Rev. Lett.
{\bf90}, 117004 (2003).

\bibitem{Hawthorn_LSCO}
D. G. Hawthorn, R. W. Hill, C. Proust, F. Ronning, M. Sutherland,
E. Boaknin, C. Lupien, M. A. Tanatar, J. Paglione, S. Wakimoto, H.
Zhang, L. Taillefer, T. Kimura, M. Nohara, H. Takagi, and N. E.
Hussey, Phys. Rev. Lett. {\bf 90}, 197004 (2003).

\bibitem{Gusynin}
V. P. Gusynin and V. A. Miransky, Euro. Phys. J. B. {\bf 37}, 363
(2004).

\bibitem{Takigawa}
M. Takigawa, M. Ichioka, and K. Machida, Physica C {\bf 404}, 375
(2004).

\bibitem{Hoffman}
J. F. Hoffman, E. W. Hudson, K. M. Lang, V. Madhavan, H. Eisaki,
S. Uchida, and J. C. Davis, Science {\bf 295}, 466 (2002).

\bibitem{Lake}
B. Lake, G. Aeppli, K. N. Clausen, D. F. McMorrow, K. Lefmann, N.
E. Hussey, N. Mangkorntong, M. Nohara, H. Takagi, T. E. Mason, and
A. Schroder, Science {\bf 291}, 1759 (2001); B. Lake, H. M.
Ronnow, N. B. Christensen, G. Appli, K. Lefmann, D. F. McMorrow,
P. Vorderwisch, P. Smeibidl, N. Mangkorntong, T. Sasagawa, H.
Nohara, H. Takagi, and T. E. Mason, Nature (London) {\bf 415}, 299
(2002).

\bibitem{Khaykovich}
B. Khaykovich, Y. S. Lee, R. W. Erwin, S.-H. Lee, S. Wakimoto, K.
J. Thomas, M. A. Kastner, and R. J. Birgeneau, Phys. Rev. B {\bf
66}, 014528 (2002).

\bibitem{Sun_PLCO}
X. F. Sun, I. Tsukada, T. Suzuki, S. Komiya, and Y. Ando, Phys.
Rev. B {\bf 72}, 104501 (2005).

\bibitem{Moler}
K. A. Moler, D. J. Baar, J. S. Urbach, R. Liang, W. N. Hardy, and
A. Kapitulnik, Phys. Rev. Lett. {\bf 73}, 2744 (1994); K. A.
Moler, D. L. Sisson, J. S. Urbach, M. R. Beasley, A. Kapitulnik,
D. J. Baar, R. Liang, and W. N. Hardy, Phys. Rev. B {\bf 55}, 3954
(1997).

\bibitem{Revaz}
B. Revaz, J.-Y. Genoud, A. Junod, K. Neumaier, A. Erb, and E.
Walker, Phys. Rev. Lett. {\bf 80}, 3364 (1998).

\bibitem{Wright}
D. A. Wright, J. P. Emerson, B. F. Woodfield, J. E. Gordon, R. A.
Fisher, and N. E. Phillips, Phys. Rev. Lett. {\bf 82}, 1550
(1999).

\bibitem{Chen}
S. J. Chen, C. F. Chang, H. L. Tsay, H. D. Yang, and J.-Y. Lin,
Phys. Rev. B {\bf 58}, R14753 (1998).

\bibitem{Brugger}
T. Brugger, T. Schreiner, G. Roth, P. Adelmann, and G. Czjzek,
Phys. Rev. Lett. {\bf 71}, 2481 (1993).

\bibitem{Hien}
N. T. Hien, V. H. M. Duijn, J. H. P. Colpa, J. J. M. Franse, and
A. A. Menovsky, Phys. Rev. B {\bf 57}, 5906 (1998).

\bibitem{Schottky}
J. P. Emerson, R. A. Fisher, N. E. Phillips, D. A. Wright, and E.
M. McCarron III, Phys. Rev. B {\bf 49}, R9256 (1994).

\bibitem{Ando_Hall}
Y. Ando, Y. Hanaki, S. Ono, T. Murayama, K. Segawa, N. Miyamoto,
and S. Komiya, Phys. Rev. B {\bf 61}, R14956 (2000); {\bf 63},
069902(E) (2001).

\bibitem{Ando_LSCO}
Y. Ando, Y. Kurita, S. Komiya, S. Ono, and K. Segawa, Phys. Rev.
Lett. {\bf 92}, 197001 (2004).

\bibitem{Abe}
Y. Abe and Y. Ando, unpublished.

\bibitem{Presland}
M. R. Presland, J. L. Tallon, R. G. Buckley, R. S. Liu, and N. E.
Flower, Physica C {\bf 176}, 95 (1991).

\bibitem{Bi2212}
P. B. Allen, X. Du, L. Mihaly, and L. Forro, Phys. Rev. B {\bf
49}, 9073 (1994).

\bibitem{Howald}
C. Howald, P. Fournier, and A. Kapitulnik, Phys. rev. B {\bf 64},
100504(R) (2001).

\bibitem{Lang}
K. M. Lang, V. Madhavan, J. E. Hoffman, E. W. Hudson, H. Eisaki,
S. Uchida, and J. C. Davis, Nature (London) {\bf 415}, 412 (2002).

\bibitem{McElroy}
K. McElroy, J. Lee, J. A. Slezak, D.-H. Lee, H. Eisaki, S. Uchida,
and J. C. Davis, Science {\bf 309}, 1048 (2005).

\bibitem{Pan}
S. H. Pan, E. W. Hudson, K. M. Lang, H. Eisaki, S. Uchida, and J.
C. Davis, Nature (London) {\bf 403}, 746 (2000).

\bibitem{Cohn}
J. L. Cohn, C. K. Lowe-Ma, and T. A. Vanderah, Phys. Rev. B {\bf
52}, R13134 (1995).

\bibitem{Sun_LCO}
X. F. Sun, J. Takeya, S. Komiya, and Y. Ando, Phys. Rev. B {\bf
67}, 104503 (2003).

\bibitem{Jin}
R. Jin, Y. Onose, Y. Tokura, D. Mandrus, P. Dai, B. C. Sales,
Phys. Rev. Lett. {\bf 91}, 146601 (2003).

\bibitem{Berggold}
K. Berggold, T. Lorenz, J. Baier, M. Kriener, D. Senff, H. Roth,
A. Severing, H. Hartmann, A. Freimuth, S. Barilo, and F. Nakamura,
Phys. Rev. B {\bf 73}, 104430 (2006).

\bibitem{Kambe}
S. Kambe, K. Okuyama, S. Ohshima, and T. Shimada, Physica C {\bf
250}, 50 (1995); X. F. Sun, X. Zhao, W. B. Wu, X. J. Fan, X. G.
Li, and H. C. Ku, Physica C {\bf 307}, 67 (1998); and references
therein.

\bibitem{Ando_Condmat}
Y. Ando, X. F. Sun, and K. Segawa, arXiv:0711.4214.

\bibitem{Hill}
R. W. Hill, C. Lupien, M. Sutherland, E. Boaknin, D. G. Hawthorn,
C. Proust, F. Ronning, L. Taillefer, R. Liang, D. A. Bonn, and W.
N. Hardy, Phys. Rev. Lett. {\bf 92}, 027001 (2004).

\bibitem{note}
The $T^3$ dependence of QP heat conductivity does not affect the
vality of fitting to the lowest-$T$ data using Eq. (\ref{T^2}) to
get $\kappa_0/T$.

\bibitem{Berman}
R. Berman, {\it Thermal Conduction in Solids} (Oxford University
Press, Oxford, 1976).

\bibitem{Smith}
M. F. Smith, J. Paglione, M. B. Walker, and L. Taillefer, Phys.
Rev. B {\bf 71}, 014506 (2005).

\bibitem{Kim}
T. K. Kim, A. A. Kordyuk, S. V. Borisenko, A. Koitzsch, M.
Knupfer, H. Berger, and J. Fink, Phys. Rev. Lett. {\bf 91}, 167002
(2003).

\bibitem{Gao}
Y. Gao, P. Pernambuco-Wise, J. E. Crow, J. O'Reilly, N. Spencer,
H. Chen, and R. E. Salomon, Phys. Rev. B {\bf 45}, 7436 (1992);
and reference therein.

\bibitem{Walton}
D. Walton, J. E. Rives, and Q. Khalid, Phys. Rev. B {\bf 8}, 1210
(1973).

\end{thebibliography}
\end{document}